\documentclass[conference]{IEEEtran}
\IEEEoverridecommandlockouts
\usepackage{cite}
\usepackage{amsmath,amssymb,amsfonts}
\usepackage{multirow}
\usepackage{array}
\usepackage{mathtools}
\usepackage{algorithmic}
\usepackage{graphicx}
\usepackage{threeparttable, tablefootnote}
\usepackage{textcomp}
\usepackage{xcolor}
\DeclarePairedDelimiter{\ceil}{\lceil}{\rceil}
\DeclarePairedDelimiter\floor{\lfloor}{\rfloor}
\newcolumntype{P}[1]{>{\centering\arraybackslash}p{#1}}
\newcolumntype{M}[1]{>{\centering\arraybackslash}m{#1}}
\def\BibTeX{{\rm B\kern-.05em{\sc i\kern-.025em b}\kern-.08em
    T\kern-.1667em\lower.7ex\hbox{E}\kern-.125emX}}
\begin{document}

\title{A $75$kb SRAM in $65$nm CMOS for In-memory Computing based Neuromorphic Image Denoising\\
}

\author{\IEEEauthorblockN{Sumon Kumar Bose}
\IEEEauthorblockA{\textit{School of Electrical and} \\
\IEEEauthorblockA{\textit{ Electronic Engineering}}
\textit{Nanyang Technological University}\\
Singapore \\
bose0003@e.ntu.edu.sg}
\and
\IEEEauthorblockN{Vivek Mohan}
\IEEEauthorblockA{\textit{School of Electrical and} \\
\IEEEauthorblockA{\textit{ Electronic Engineering}}
\textit{Nanyang Technological University}\\
Singapore \\
vivekmoh001@e.ntu.edu.sg}
\and
\IEEEauthorblockN{Arindam Basu}
\IEEEauthorblockA{\textit{School of Electrical and} \\
\IEEEauthorblockA{\textit{ Electronic Engineering}}
\textit{Nanyang Technological University}\\
Singapore\\
arindam.basu@ntu.edu.sg}
}


\maketitle
\begin{abstract}
This paper presents an in-memory computing (IMC) architecture for image denoising. The proposed SRAM based in-memory processing framework works in tandem with approximate computing on a binary image generated from neuromorphic vision sensors. Implemented in TSMC $65$nm process, the proposed architecture enables $\approx 2000X$ energy savings ($\approx222X$ from IMC) compared to a digital implementation when tested with the video recordings from a DAVIS sensor and achieves a peak throughput of $1.25-1.66$ frames/$\mu s$.
\end{abstract}


\section{Introduction}
Bio-inspired neuromorphic vision sensors (NVS)~\cite{Posch2014}\cite{Berner2013} have gained traction among the researchers due to its low bandwidth and energy requirement, as well as recent commercial availability~\cite{Vision}. Unlike a traditional frame-based camera, NVS detects any changes of contrast in a pixel and outputs an event corresponding to the (x,y) coordinates of that pixel which is also known as Address Event Representation (AER)~\cite{Boahen2004}. Hence, AER contains less superfluous information due to its asynchronous event-based encoding and finds application in robotics~\cite{Delbruck2013}, environment monitoring, traffic surveillance~\cite{Litzenberger2006} and object tracking~\cite{1542972} in the scene. However, the raw image data is corrupted with noise and its removal from the image is one of the most important pre-processing tasks for region proposal, object tracking and classification~\cite{Padala2018}. While earlier approaches use an event based denoise method termed the nearest neighbour filter (NN-filt)\cite{nnfilt}, recent hybrid frame-event approaches using median filtering outperformed NN-filt in terms of performance, memory requirement and computes~\cite{Jyotibdha_EBBIOT}. However, traditional Von Neumann architecture is still a bottleneck in terms of latency and energy dissipation for hardware implementation of neuromorphic processing\cite{jetcas_review}.
\begin{figure}[t]
\centering
\includegraphics[scale=0.6]{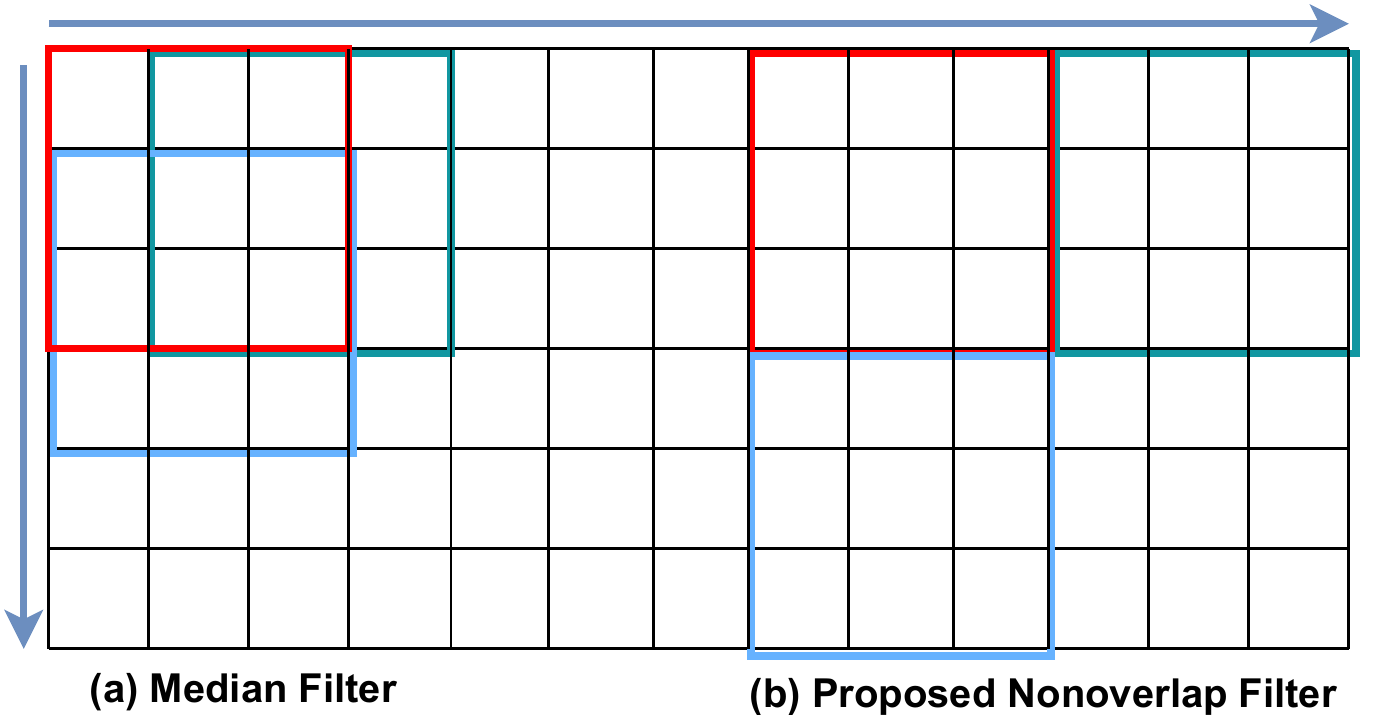}
\caption{(a) Conventional median filter using a $3\times3$ kernel and stride = 1 (b) Proposed NOMF using a $3\times3$ kernel and stride = 3 for image denoising. Approximation due to NOMF reduces the memory read, write energy and its architecture enables in-memory computing.}
\label{fig1}
\end{figure}

To address this, in-memory computing(IMC) paradigm is proposed where processing is performed inside the memory and shows unprecedented performance benefits compared to its Von Neumann counterpart. IMC not only enables highly parallel processing due to its simultaneous access of multiple cells but also gets rid of the energy consumption of data transfer from memory to processor and vice versa~\cite{Biswas2018_thesis}. Several works on IMC are shown to be effective, such as~\cite{7875410} proposed $6$-T SRAM based linear classifier using current summation and achieved $13$x energy savings on MNIST dataset compared to the digital implementation. Similarly,~\cite{Biswas2018} implemented $10$T-SRAM based binary-weighted Convolutional Neural Networks (CNN) leveraging charge distribution and attained $16$x energy benefit for MNIST dataset. \emph{While most of the efforts on IMC are shown for the post-processing of the image, in this paper, we use IMC for efficient denoising of the event based binary image (EBBI) since this method is shown to outperform pure event based ones\cite{Jyotibdha_EBBIOT}.}  

Approximate computing is another avenue for energy reduction in an application like pattern recognition or multimedia processing where slight degradation in the calculation does not affect the final outcome or the output quality remains its acceptable range. Approximation in the calculation can be introduced to the circuit~\cite{Lu2004}~\cite{approx_float}, software~\cite{Bose_2019} or system level~\cite{Raha2018xxx}. Since a slight change of object boundary has a little impact on region proposal, objects tracking or classification performance, we propose to use approximate computing while filtering of an image frame. The details of the algorithm and VLSI implementation are presented in the following sections.

\section{Overview: Median Filter Algorithm}
\label{brief}
A median filter is a nonlinear filter that replaces the center pixel of an $n\times n$ kernel by the median value of $n^2$ pixels associated with the kernel. The output of median filter at (i, j) location can be presented as Eq.(\ref{eqn1}) where i, j $\subset \mathbb{Z}^+$.
  \begin{equation}
 \begin{split}
P_{mf}(i,j) = &median(\{P(i+k,j+l) ~|~k,l \subset \mathbb{Z} \\& \text{and}~\in \{-\frac{n-1}{2},\cdots,\frac{n-1}{2}\})
\label{eqn1}
\end{split}
\end{equation}
Implementation of the median filter for a grayscale image involves sorting the pixel values. On the contrary, carrying out the median filtering for a binary image  is simple and requires a counter which adds up the number of occurrence of ``1" for an $n\times n$ patch and assign ``1" for the middle pixel if the number of ``1"s is higher than that of ``0" and vice versa. The whole operation can be shown as 
\begin{align}
 P_{mf}(i,j)&=
\begin{dcases}
    1,& \text{if $\Sigma P(i+k,j+l) \geq \ceil*{\frac{n^2}{2}}$ }\\
    0,& \text{otherwise} \label{eqn2anew}
\end{dcases}
\end{align}

In a traditional median filter, an $n\times n$ kernel convolves over the image in an overlap fashion where the stride, $s=1$ as shown in Fig.~\ref{fig1}(a). Hence, fetching and summing up bit by bit for the binary image, followed by comparison in the processing unit and a write operation in the memory demand $2n^2$+1 clock cycles and associated energy for each pixel. However, since the adjacent pixels of an image have similar characteristics, we can apply the decision of an $n\times n$ kernel to all the $n^2$ pixels instead of the center one. This is equivalent to having stride $s=n$ (Fig. \ref{fig1}(b)) resulting in \textbf{non-overlap median filter (NOMF)} that we use in this work. While the proposed approach changes the object boundary slightly (marginal effect on tracking as shown later), it reduces the processing and memory read access time by a factor of $n^2$ and enables the same memory to be utilized to store the filtered image. It also enables IMC based denoise as shown next. However, NOMF approach does not reduce the memory write cycles and energy. Table \ref{tab2} captures usage of the resources in both approaches for an image of size $W\times H$.
\section{In-memory Denoise: Hardware implementation}
\label{hw}
\subsection{Architecture}
\begin{figure}[t]
\centering
\includegraphics[scale=0.52]{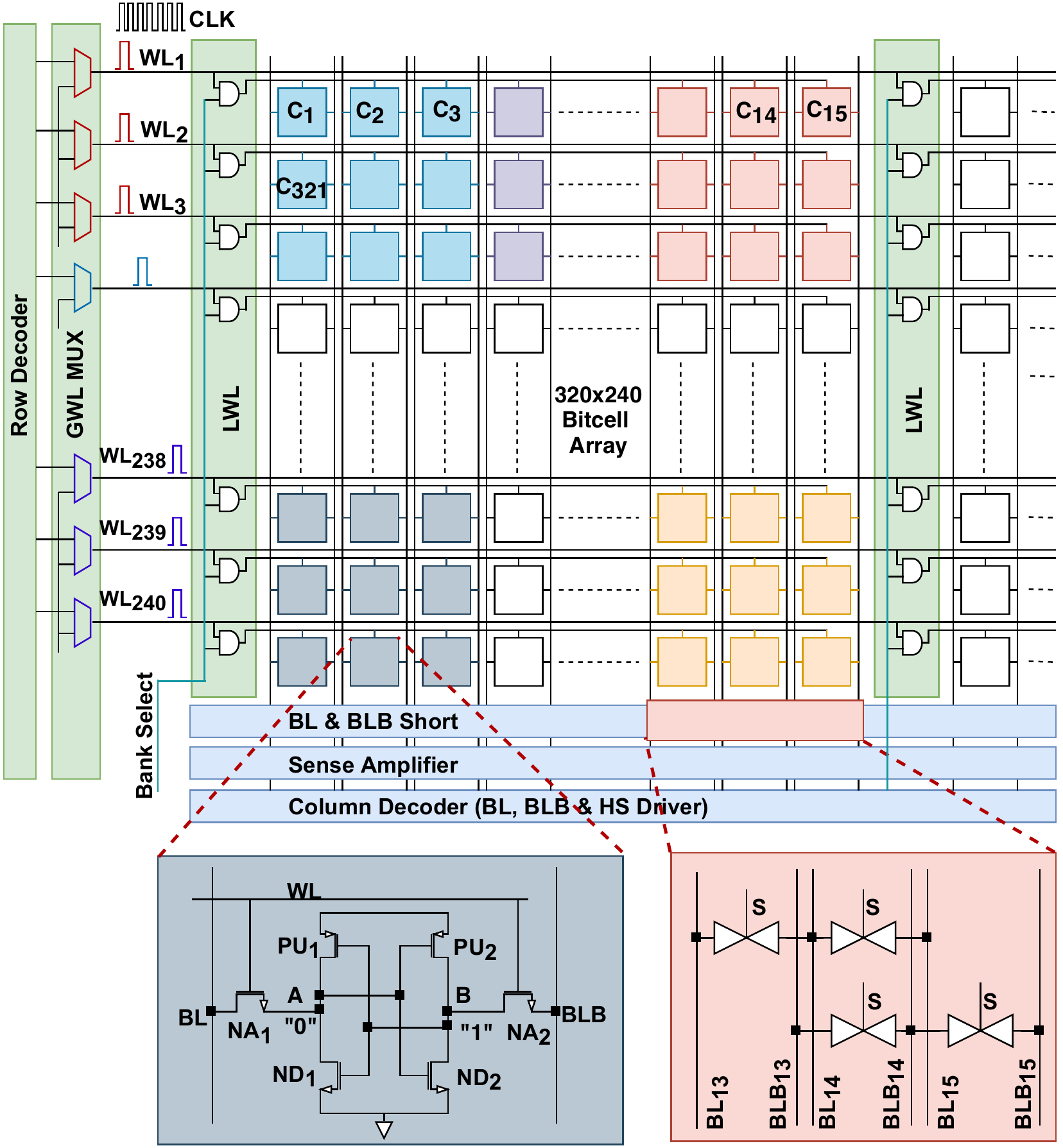}
\caption{Architecture of a $320\times240$ bitcell array for noise removal from a binary image. In filter mode for $n=3$, three consecutive word-lines (WL) are enabled together to discharge bit-line (BL) and bit-line bar (BLB) simultaneously. BLs and BLBs of three successive columns are connected together separately using transmission gates to implement a $3\times3$ kernel. The IMC architecture enables highly parallel noise filtering of $320\times3$ cells in two clock cycles.}
\label{fig2}
\end{figure}

Figure \ref{fig2} shows an architecture of a $320\times240$ SRAM array for image denoising (QVGA or lower resolution) applicable to NVS such as~\cite{5648367}~\cite{Brandli:200837}. It operates in two modes \textbf{(a)} normal read and write mode \textbf{(b)} filter mode. Unlike a conventional SRAM write, NVS does not allow to write all the bits of a byte or a word simultaneously since this memory is targeted for event-based cameras and events are not contiguous. Therefore, a single bit writing circuitry is implemented in normal write mode. In order to reduce the dynamic bit-line power consumption~\cite{Chandrakasan:1995:LPD:560639}, the whole memory is divided into $22$ banks having $15\times240$ cells in each bank except the last one. In filter mode, the kernel can be configured as either a $3\times3$ or a $5\times5$ (enabling $n$ successive WLs and connecting $n-1$ consecutive BLs and BLBs separately, $n\in \{3,5\}$)  patch. To have almost the same delay of WL signal for each cell of a kernel, $15$ columns are selected for each bank. In normal SRAM write mode, global (GWL) and local word-line (LWL) blocks enable one of the word-lines (WL), and column decoder writes the data and its complement on the bit-line (BL) and bit-line bar (BLB) respectively. The rest of the BLs and BLBs are charged to VDD by the half select (HS) driver to mitigate the read disturb issue of the half-selected cells in the selected bank (cells are selected along row but not selected along column).

During writing a memory cell, one of the lines (either BL or BLB) is driven to $0$ V and another line is connected to VDD. The line, connected to $0$ V, initiates the bit-flip process in an SRAM cell. For instance, 6T SRAM cell in the left inset of Fig. \ref{fig2} stores ``0" and in order to write ``1" in the cell, BL and BLB are connected to VDD and $0$ V respectively. Once WL is asserted, the strength of $\text{PU}_2$ and $\text{NA}_2$ decides the bit-flip in the cell. If $\text{NA}_2$ has higher strength than $\text{PU}_2$, it will write ``1" in the cell. However, the writing operation can happen even when the BL is connected to lower potential than VDD. In that case, strength of  $\text{NA}_2$ transistor has to be increased further. In read mode, BL and BLB are charged to VDD, and when the WL signal is asserted, either of the lines starts discharging depending on the value stored in the cell.

\subsection{Implementation of NOMF}
We follow the steps of an SRAM cell read and bit-flip to implement the NOMF for noise removal in the memory. BLs and BLBs of the $3\times3$ cells are connected separately employing transmission gates which is shown in the right inset of Fig. \ref{fig2}. Throughout the filter operation, the signal $\text{S}$ is kept high. The resistance of the transmission gate, $R_{tg}$ is chosen such that the following criterion is met:
\begin{equation}
R_{tg}C_{BL}<<C_{BL}\cdot\frac{VDD}{i_s}
\label{eqn3}
\end{equation}
where $i_s$ denotes the discharging current of each SRAM cell and $C_{BL}$ is a combination of the metal routing capacitor of  BL or BLB, and diffusion capacitor of $240$ access transistors, $\text{NA}_1$ or $\text{NA}_2$. From post-layout simulation after parasitic extraction, $C_{BL}$ $\approx 140$fF. The condition in Eq. (\ref{eqn3}) is maintained so that the discharge profiles of the three BLs of a $3\times3$ kernel follow each other with minimal delay and the same is applicable for the BLBs discharge. The proposed IMC architecture takes two clock cycles to filter the noise from a $n\times n$ patch. In the first cycle, $n$ BLs and BLBs are charged to VDD. $n$ successive WLs are asserted in the next cycle, which enables $n^2$ ($n\times n$) cells to discharge BLs and BLBs simultaneously. Since there will be a difference of BL and BLB discharge current due to the different number of ``0"s and ``1"s in a $n\times n$ kernel, one of the lines will discharge and reach $0$ V faster. This configuration of BL and BLB is similar to write mode and it will flip the minority pixels in the kernel. If the number of ``0"s is less than the number of ``1"s in a kernel, we refer ``0" as minority pixel in that patch and vice versa. In filter mode, we keep all the bank select signals high to activate highly parallel processing in the memory and it filters $320\times3$ cells in one pass. We repeat this procedure until all the rows are filtered.

Intuitively, the $n\times n$ kernel can be thought of as a circuit where two latches of different strength and stored values are connected to BL and BLB. Their strengths are determined by the number of ``0"s and ``1"s stored in the kernel. Whoever wins in discharging BL or BLB faster, imposes its stored value on the other.

\begin{table}[t]
\centering
\caption{Comparison of different filters for an image of size, $\text{D}$=$\text{W}\times\text{H}$}
 \begin{threeparttable}
    \begin{tabular} {|M{1.57cm}|M{0.60cm}|M{0.94cm}|M{0.94cm}|M{1.2cm}|M{0.65cm}|}\hline
          & Input &\# memory read &\# memory write &\# operations &\# Bits \\ \hline
        NN-filt &Events &$\beta_t\gamma n^2\text{D}\tnote{}$ &$\beta_t\gamma \text{D}$  &$\gamma n^2\text{D}$ &$\beta_t \text{D}$\\ \hline
        Median Filter &EBBI	 &$n^2\text{D}$  &$\text{D}$  &$n^2\text{D}$ &$2\text{D}$\\ \hline
        NOMF &EBBI&$\text{D}$  &$\text{D}$ &$\text{D}$ &$\text{D}$\\ \hline
        NOMF+IMC &EBBI &$\text{D}/n$  &$\alpha\text{D}$ &$0$ &$\text{D}$\\ \hline
    \end{tabular}
\label{tab2}
\begin{tablenotes}
   \item[] $\beta_t=16$, $\gamma\approx15\%$, $\text{D}=\text{HW}$, $\alpha\approx3.6\%$, $n\in \{3,5\}$.  
  \end{tablenotes}
  \end{threeparttable}
\end{table}

\begin{figure*}[t]
\centering
\includegraphics[scale=0.6]{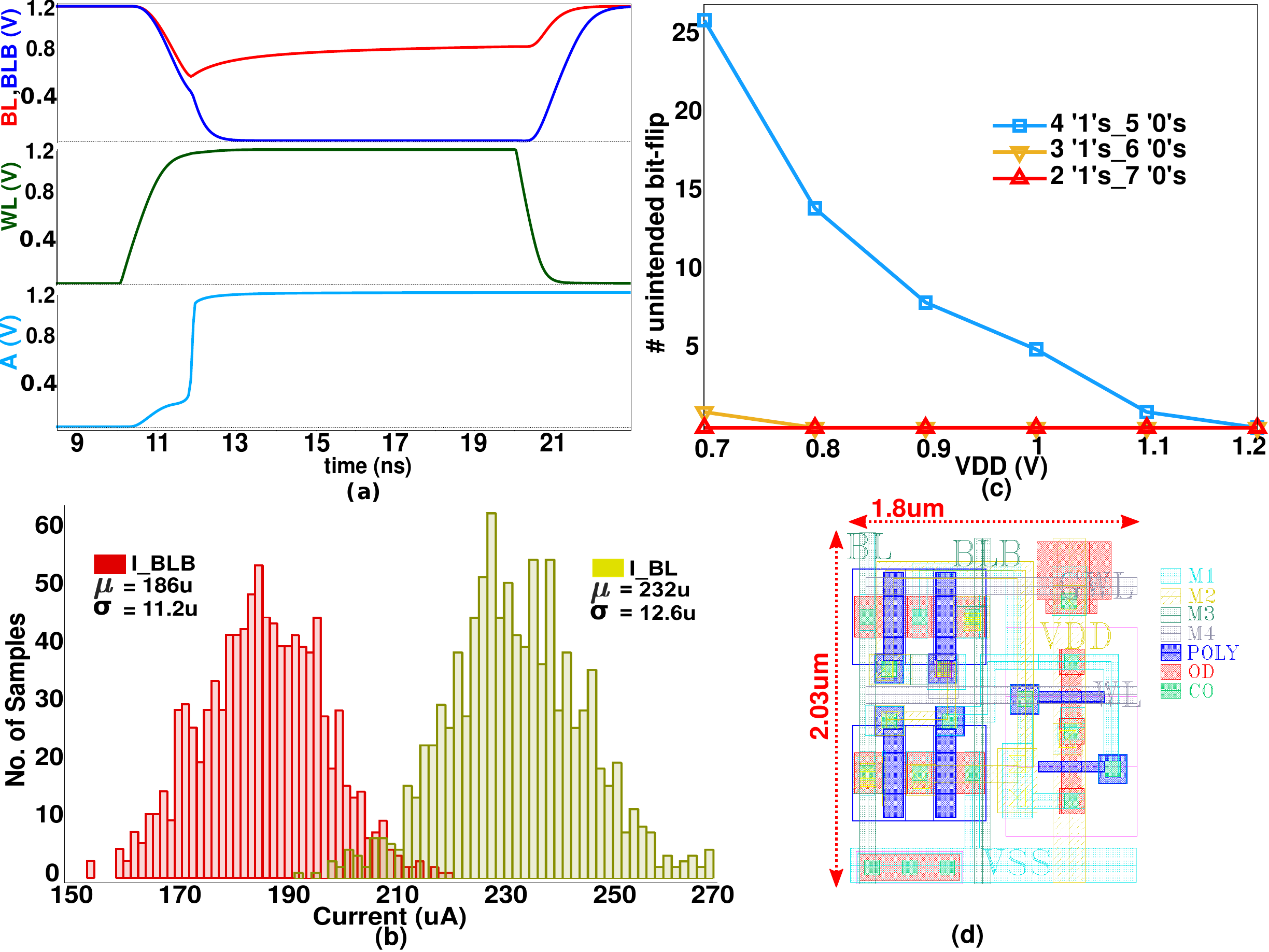}
\caption{(a) Bit-flip of an SRAM memory cell in a $3\times3$ kernel. Since the number of ``1" is higher than that of ``0", BLB gets discharged faster and the stored value flips at node A. (b) $1000$ points Monte-Carlo DC simulation of BL and BLB discharging current at VDD=$1$V, and $4``1"\text{s}\_5``0"\text{s}$ scenario (worst-case). (c) $200$ points Monte-Carlo simulation: unintended bit-flip due to the mismatches across VDD and different number of ``1"s and ``0" in the kernel. (d) Unit SRAM cell layout.}
\label{fig3}
\end{figure*}

The voltage difference between BL and BLB at any instant of time, $t$, is represented as
\begin{equation}
\Delta V=\bigg(\frac{\Sigma i_0}{C_{BL}}-\frac{\Sigma i_1}{C_{BLB}}\Bigg)t
\label{eqn4}
\end{equation}
Where $\Sigma i_0$ and $\Sigma i_1$ represent the discharging current of BL and BLB due to the stored ``0"s and ``1"s in the kernel respectively. In the best-case scenario, all the bits in the kernel are either ``0" or ``1" and bit-flip does not happen. In contrast, the kernel takes the longest time to decide and flips the minority pixels when the difference between the number of ``0"s and ``1" is one. However, due to the discharging current and capacitor mismatch, majority pixels in a kernel may flip in the worst-case scenario. The unintended bit-flips due to the mismatch reduces the object boundary when the majority pixel is ``1" and inserts new object in the frame in the opposite scenario ($\floor*{\frac{n^2}{2}}$ ``0"s and $\ceil*{\frac{n^2}{2}}$ ``1"s). However, the probability of  $\floor*{\frac{n^2}{2}}$ noise pixels appearing inside the faulty kernel is negligible. Nevertheless, to mitigate the mismatch effects, width and length of $\text{NA}_1$, $\text{NA}_2$, $\text{ND}_1$, and $\text{ND}_2$ are increase by a factor of $2$ from its minimum value supported by the process and low VT devices are used. We run $200$ trials of Monte-Carlo simulation initializing the kernel with four ``1"s and five ``0" and do not observe any unintentional bit-flip in the $3\times3$ kernel (see Fig.~\ref{fig3}(c)) at VDD=$1.2$V. Even though the usage of low VT devices increases the leakage power, we can shut down the memory once processing is done.
\subsection{Performance}
The proposed approach has several major advantages \textbf{a)} it reduces the dynamic BL power consumption during SRAM read operation. BLs and BLBs are required to charge once to read $n$ (3 or 5) cells along the column compared to the conventional approach where the requirement is $n$ times. \textbf{b)} It does not require any sense amplifier to sense the BL and BLB voltage difference. The $n\times n$ kernel inherently acts as a sense amplifier and takes the decision. \textbf{c)} It does not consume any dynamic BL power during write operation since the discharges of BL and BLB are related to the read operation. \textbf{d)} Minimal energy is required to flip the minority pixels (only noise and boundary pixels of an object).

Table \ref{tab2} compares the proposed NOMF implemented using IMC with the state-of-the-art denoising techniques. The nearest neighbour filter (NN-filt)~\cite{Gonzalez:2006:DIP:1076432} stores and updates the timestamp of an incoming event using $\beta_t$ ($\beta_t$=16) bit per timestamp~\cite{Jyotibdha_EBBIOT}. Whereas other techniques process event based-binary image (EBBI) frame. $\gamma$ represents the average number of events ($\approx15\%$) during the frame duration (single pixel can be fired multiple times). As discussed earlier, IMC reduces the number of memory read by a factor of $n$. $\alpha$ in the third column represents the fraction of the pixels that need to be flipped for the filter implementation (only noise and boundary pixels of the objects). We observed that the average value of $\alpha$  is $0.036$ for $15000$ image frames. Also, the proposed IMC approach does not require any addition or comparison.

\section{results}
\label{re}
The circuit has been designed in $65$nm CMOS refer to unit SRAM cell layout picture in Fig.~\ref{fig3}(d). We initialize one of the $3\times3$ kernels of the $320\times240$ memory array with four ``0"s and five ``1" to simulate the NOMF in SPICE. Fig. \ref{fig3}(a) captures the transient behavior of different nodes of the kernel. When WL goes low, BL and BLB are charged to VDD. Initially node A stores ``0" and when WL is made high, BL and BLB start discharging. Since the number of ``1" is higher than that of ``0" in the kernel, BLB gets discharged faster and the minority cells flip its stored value. $1000$ points Monte-Carlo DC simulation of BL and BLB discharging current at the worst-case scenario and $1$V is shown in Fig.\ref{fig3}(b). The overlap region in the histogram is responsible for the unintended bit-flips at 1V. However, it is seen in MATLAB simulation using the dataset described below that the probability of appearing four noisy pixels in a kernel is $0.0006$. Fig. \ref{fig3}(c) shows $200$ points Monte-Carlo simulation of unintended bit-flips across VDD. It can be seen that at $1.2$V, unintended bit-flip does not happen. However, due to lower overdrive, and mismatches, $2.5\%$ unintended bit-flips occur at $1$V and the worst-case scenario (Overall probability=$0.025\times0.0006$).

In order to validate the proposed NOMF and compare with prior work, we use the same dataset as used in \cite{Jyotibdha_EBBIOT} for a fair comparison. The dataset comprises more than 1 hour of traffic scene recordings with different objects such as cars, buses, trucks, bikes, humans along with the background noise. More details are available in \cite{Jyotibdha_EBBIOT}.

\begin{figure*}[t]
\centering
\includegraphics[scale=0.6]{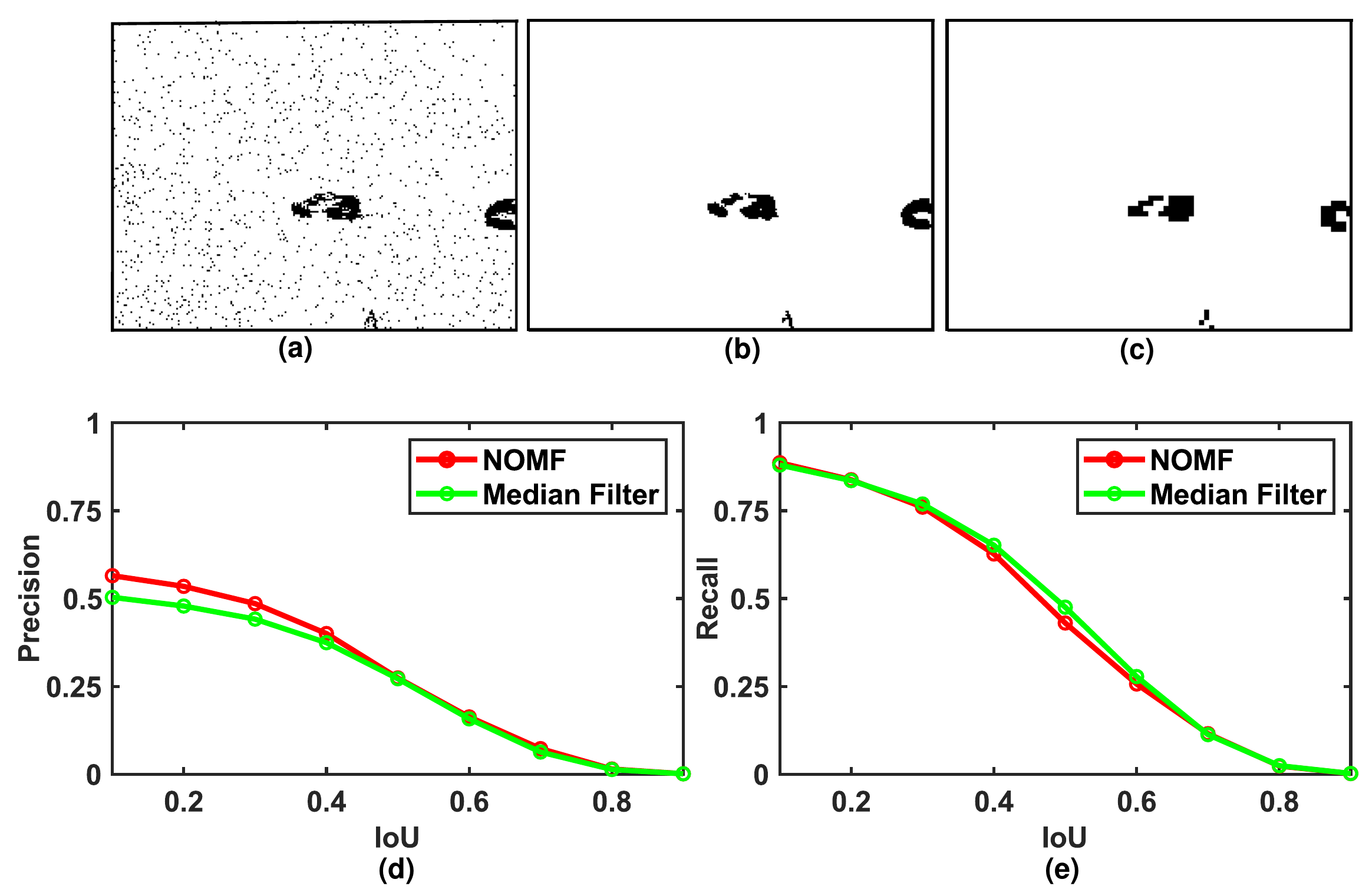}
\caption{(a) raw binary frame (b) output frame of median filter and (c) proposed NOMF using a $3\times3$ kernel. (d) Precision and (e) recall of an overlap-based tracker (OT)~\cite{Jyotibdha_EBBIOT} using both filtered images for different IoU values.}
\label{fig4}
\end{figure*}

Fig. \ref{fig4}(b)-(c) show the MATLAB simulation of the median filter and proposed NOMF using a $3\times3$ kernel on the binary raw image. In term of noise removal, both filters show similar performance. We also evaluate the performance - recall and precision of an overlap-base tracker (OT)~\cite{Jyotibdha_EBBIOT} using both filtered images for different IoU values as shown in Fig.~\ref{fig4}(d)-(e). $\text{IoU=}(A_{GTB}\cap A_{PB})/(A_{GTB}\cup A_{PB})$, where $A_{GTB}$ and $A_{PB}$ denote the area of manually annotated ground truth and region proposed by the OT encapsulating an object respectively. If the IoU of a proposed region is greater than a threshold, the region is assumed to be true positive region. Precision (true positive regions/ total proposed regions) and recall (true positive regions/ total ground truth regions) of the OT are calculated using all the output frames from both filters, and the performance is comparable as shown in Fig.\ref{fig4}.

Table \ref{tab4} compares the performance of the proposed NOMF implemented using IMC with spatiotemporal~\cite{7168735} and fully-digital median filter that is synthesized in the same process for fair comparison. The spatiotemporal filter works on the continuous events from the NVS whereas proposed NOMF and fully-digital implementation process event-based binary image following \cite{Jyotibdha_EBBIOT}. Latency and energy are estimated at $200$MHz on the post-layout netlist. The synergy between the approximate computing and IMC reduces the execution time to $0.8\mu s$/frame and enables $\approx 2000X$ energy saving compared to the digital counterpart where the contribution of approximation and IMC are $\approx9X$ and $\approx222X$ respectively.
\begin{table}[t]
\centering
\caption{Comparison with different filter implementations}
    \begin{tabular} {|M{1.9cm}|M{0.7cm}|M{1cm}|M{1.25cm}|M{1.12cm}|}\hline
          &Process &Area/Cell ($\mu m^2$) &Latency($ns$) /bit &Energy(pJ) /bit \\ \hline
        Spatio-temporal
        Filter~\cite{7168735} &180nm &400  &10  &20\\ \hline
        Median Filter	 &65nm &4.89  &95  &228\\ \hline
    
       Proposed NOMF+IMC &65nm &3.65  &0.01 &0.11\\ \hline
    \end{tabular}
\label{tab4}
\end{table}

\begin{table}[h!]
\centering
\caption{Comparison of different published IMC works}
    \begin{tabular} {|M{1.8cm}|M{1.3cm}|M{1cm}|M{1cm}|M{1cm}|}\hline
          &This Work &~\cite{Biswas2018} &~\cite{Kang2018} &~\cite{8662392} \\ \hline
        Technology	 &65nm  &65nm  &65nm &55nm\\ \hline
        Algorithm &Filter  &CNN &k-NN &CNN\\ \hline
         SRAM size &75kb &16kb &128kb &3.75kb\\ \hline
        Throughput (GOPS) &85.3-153  &8 &10.2 &-\\\hline
        Energy Efficiency(TOPS/W) &11.3-20  &14.7-40.3 &1.94 &18.37-72.1\\\hline
    \end{tabular}
\label{tab5}
\end{table}
Table \ref{tab5} compares the proposed approach with the recently published IMC works and demonstrates an order of magnitude improvement in throughput due to the highly parallel processing. Assuming $n^2-1$ operations (addition) for the calculation of a $n\times n$ kernel, the energy efficiency is comparable with other state of the art.
\section*{Conclusion}
 In this work, we present an approximate and in-memory computing framework for binary image denoising. The proposed approach is tested with the binary image frames from a DAVIS sensor setup and achieves $\approx2000X$ energy saving compared to conventional Von Neumann digital approaches. The massively parallel architecture
 reduces the processing time to $0.6\mu s$ per frame and provides enough time for the subsequent processing stages.

\bibliographystyle{IEEEtran}
\bibliography{imc.bib}

\end{document}